\documentclass[11pt,a4paper,twoside]{article}
\usepackage{graphicx}
\usepackage{color}
\begin{document}
\baselineskip=0.20in
\vspace{20mm}
\baselineskip=0.30in
{\bf \LARGE
\begin{center}
Approximate bound state solutions of the deformed Woods-Saxon potential using asymptotic iteration method
\end{center}}
\vspace{4mm}
\begin{center}
{\Large {\bf Babatunde J. Falaye}}\footnote{\scriptsize E-mail:~ fbjames11@physicist.net} \\
\small
\vspace{2mm}
{\it Theoretical Physics Section, Department of Physics\\
 University of Ilorin,  P. M. B. 1515, Ilorin, Nigeria.}
\vspace{4mm}
\end{center}
\begin{center}
{\Large {\bf Majid Hamzavi}}\footnote{\scriptsize E-mail:~ majid.hamzavi@gmail.com} \\
\small
\vspace{2mm}
{\it Department of Basic Sciences, Shahrood Branch,\\
 Islamic Azad University, Shahrood, Iran}
 \end{center}
\vspace{4mm}
\begin{center}
{\Large {\bf Sameer M. Ikhdair}}\footnote{\scriptsize E-mail:~ sikhdair@neu.edu.tr} \\
\small
\vspace{2mm}
{\it Physics Department, Near East University, \\
922022 Nicosia, North Cyprus, Mersin 10, Turkey}
\end{center}
\vspace{12mm}
\noindent
\begin{abstract}
\noindent
By using the Pekeris approximation, the Schr$\ddot{o}$dinger equation is approximately solved for the nuclear deformed Woods-Saxon potential within the framework of the asymptotic iteration method.  The energy levels are worked out and the corresponding normalized eigenfunctions are obtained in terms of hypergeometric function. 
\end{abstract}


iteration method.

{\bf PACs No.} 03.65.Pm; 03.65.Ge; 03.65.-w; 03.65.Fd; 02.30.Gp

\section{Introduction}
The deformed Woods-Saxon (dWS) potential is a short range potential and widely used in nuclear, particle, atomic, condensed matter and chemical physics [1-7]. This potential is reasonable for nuclear shell models and used to represent the distribution of nuclear densities. The dWS and spin-orbit interaction are important and applicable to deformed nuclei $\cite{BJ8}$ and to strongly deformed nuclides $\cite{BJ9}$. The dWS potential parameterization at large deformations for plutonium $^{237, 239, 241}Pu$  odd isotopes was analyzed $\cite{BJ10}$. The structure of single-particle states in the second minima of $^{237, 239, 241}Pu$ has been calculated with an exactly dWS potential. The Nuclear shape was parameterized. 

The parameterization of the spin-orbit part of the potential was obtained in the region corresponding to large deformations (second minima) depending only on the nuclear surface area. The spin-orbit interaction of a particle in a non-central self consistent field of the WS type potential was investigated for light nuclei and the scheme of single-particle states has been found for mass number $A_o=10$ and $25$ $\cite{BJ8}$. Two parameters of the spin-orbit part of the dWS potential, namely the strength parameter and radius parameter were adjusted to reproduce the spins for the values of the nuclear deformation parameters $\cite{BJ11}$.

Badalov et al. investigated Woods-Saxon potential in the framework of Schr$\ddot{o}$dinger and Klein-Gordon equations by means of Nikiforov-Uvarov method $\cite{BJ12, BJ13}$. In our recent works $\cite{BJ14, BJ15}$, we have studied the relativistic Duffin-Kemmer-Petiau and the Dirac equation for dWS potential. In addition, we have also obtained the bound state solutions of the $PT-/non-PT$-symmetric and non-Hermitian modified Woods-Saxon potential with the real and complex-valued energy levels $\cite{BJS1}$. 

In this paper we will study Schr$\ddot{o}$dinger equation with dWS potential for any arbitrary orbital quantum number $\ell$. We obtain the analytical expressions for the energy levels and wave functions in closed form. Therefore, this work is arranged as follows: In section $2$, a brief introduction to asymptotic iteration method (AIM) is given. In section $3$, the Schr$\ddot{o}$dinger equation with dWS potential is solved. Finally, results and conclusions are presented in section $4$.

\section{The Asymptotic Iteration Method}
We briefly outline the AIM here; the details can be found in references $\cite{BJ16, BJ17}$.
\subsection{Energy eigenvalues}
AIM is proposed to solve the homogenous linear second-order differential equation of the form
\begin{equation}
y_n''(x)=\lambda_o(x)y_n'(x)+s_o(x)y_n(x),
\label{E1}
\end{equation}
where $\lambda_o(x)\neq0$ and the prime denotes the derivative with respect to $x$, the extral parameter $n$ is thought as a radial quantum number. The variables, $s_o(x)$ and  $\lambda_o(x)$ are sufficiently differentiable. To find a general solution to this equation, we differentiate equation (\ref{E1}) with respect to $x$ as
\begin{equation}
y_n'''(x)=\lambda_1(x)y_n'(x)+s_1(x)y_n(x),
\label{E2}
\end{equation}
where
\begin{eqnarray}
\lambda_1(x)&=&\lambda_o'(x)+s_o(x)+\lambda_o^2(x),\nonumber\\
s_1(x)&=&s_o'(x)+s_o(x)\lambda_o(x),
\label{E3}
\end{eqnarray}
and the second derivative of equation (\ref{E1}) is obtained as
\begin{equation}
y_n''''(x)=\lambda_2(x)y_n'(x)+s_2(x)y_n(x),
\label{E4}
\end{equation}
where
\begin{eqnarray}		
\lambda_2(x)&=&\lambda_1'(x)+s_1(x)+\lambda_o(x)\lambda_1(x),\nonumber\\
	 s_2(x)&=&s_1'(x)+s_o(x)\lambda_1(x).
\label{E5}
\end{eqnarray}
Equation (\ref{E1}) can be iterated up to $(k+1)th$ and $(k+2)th$ derivatives, $k=1,2,3...$ Therefore we have
\begin{eqnarray}
y_n^{(k+1)}(x)&=&\lambda_{k-1}(x)y_n'(x)+s_{k-1}(x)y_n(x),\nonumber\\
y_n^{(k+2)}(x)&=&\lambda_{k}(x)y_n'(x)+s_{k}(x)y_n(x),
\label{E6}
\end{eqnarray}
where
\begin{eqnarray}
\lambda_k(x)&=&\lambda_{k-1}'(x)+s_{k-1}(x)+\lambda_o(x)\lambda_{k-1}(x),\nonumber\\
s_k(x)&=&s_{k-1}'(x)+s_o(x)\lambda_{k-1}(x).
\label{E7}
\end{eqnarray}
From the ratio of the (k+2)th and (k+1)th derivatives, we obtain
\begin{equation}
\frac{d}{dx}ln\left[y_n^{k+1}(x)\right]=\frac{y_n^{(k+2)}(x)}{y_n^{(k+1)}(x)}=\frac{\lambda_k(x)\left[y_n'(x)+\frac{s_k(x)}{\lambda_k(x)}y_n(x)\right]}{\lambda_{k-1}(x)\left[y_n'(x)+\frac{s_{k-1}(x)}{\lambda_{k-1}(x)}y_n(x)\right]},
\label{E8}
\end{equation}
if $k>0$, for sufficiently large $k$, we obtain $\alpha$ values from
\begin{equation}
\frac{s_k(x)}{\lambda_k(x)}=\frac{s_{k-1}(x)}{\lambda_{k-1}(x)}=\alpha(x),
\label{E9}
\end{equation}
with quantization condition
\begin{equation}
\delta_k(x)=
\left|
\begin{array}{lr}     
\lambda_k(x)&s_k(x) \\      
  \lambda_{k-1}(x)&s_{k-1}(x)
  \end{array}
  \right|=0\ \ ,\ \  \ k=1, 2, 3....
\label{E10}
  \end{equation}
Then equation (\ref{E8}) reduces to 
\begin{equation}
\frac{d}{dx}ln\left[y_n^{(k+1)}(x)\right]=\frac{\lambda_k(x)}{\lambda_{k-1}(x)},
\label{E11}
\end{equation}
which yields the general solution of Eq. (\ref{E1})
\begin{equation} y_n(x)=\exp\left(-\int^x\alpha(x')dx'\right)\left[C_2+C_1\int^x\exp\left(\int^{x'}\left[\lambda_o(x'')+2\alpha(x'')\right]dx''\right)dx'\right].
\label{E12}
\end{equation}
For a given potential, the idea is to convert the radial Schr$\ddot{o}$dinger equation to the form of equation (\ref{E1}). Then $\lambda_o(x)$ and $s_o(x)$ are determine and $s_k(x)$ and $\lambda_k(x)$ parameters are calculated by the recurrence relations given by equation (\ref{E7}). The energy eigenvalues are then obtained by the condition given by equation (\ref{E10}) if the problem is exactly solvable.
\subsection{Energy eigenfunction}
Suppose we wish to solve the radial Schr$\ddot{o}$dinger equation for which the homogenous linear second-order differential equation takes the following general form
\begin{equation}
y''(x)=2\left(\frac{tx^{N+1}}{1-bx^{N+2}}-\frac{m+1}{x}\right)y'(x)-\frac{Wx^N}{1-bx^{N+2}}.
\label{E13}
\end{equation}
The exact solution $y_n(x)$ can be expressed as $\cite{BJ17}$
\begin{equation}
y_n(x)=(-1)^nC_2(N+2)^n(\sigma)_{_n}{_2F_1(-n,\rho+n;\sigma;bx^{N+2})},
\label{E14}
\end{equation}
where the following notations has been used
\begin{equation}
(\sigma)_{_n}=\frac{\Gamma{(\sigma+n)}}{\Gamma{(\sigma)}}\ \ ,\ \ \sigma=\frac{2m+N+3}{N+2}\ \ and\ \ \rho=\frac{(2m+1)b+2t}{(N+2)b}.
\label{E15}
\end{equation}
\section{Any $\ell$-state solutions}
The deformed Woods-Saxon potential we investigate in this study is defined as \cite{BJ14, BJ15, BJS2}
\begin{equation}
V(r)=-\frac{V_o}{1+q\exp{(\frac{r-R}{a})}},\ \ \ \ R=r_oA_o^{1/3},\ \ \ \ V_o=(40.5+0.13A_o)MeV,\ \ \ \ R>>a,\ \ \ \ q>0,
\label{E16}
\end{equation}
where $V_o$ is the depth of potential, $q$ is a real parameter which determines the shape (deformation) of the potential, $a$ is the diffuseness of the nuclear surface, $R$ is the width of the potential, $A_o$ is the atomic mass number of target nucleus and $r_o$ is radius parameter. By inserting this potential into the Schr$\ddot{o}$dinger equation \cite{BJ16, BJ17} as
\begin{equation}
\left(-\frac{\hbar^2}{2\mu}\left[\frac{1}{r^2}\frac{\partial}{\partial r}r^2\frac{\partial}{\partial r}+\frac{1}{r^2\sin \theta}\frac{\partial}{\partial\theta}\left(\sin\theta\frac{\partial}{\partial\theta}\right)+\frac{1}{r^2\sin^2\theta}\frac{\partial^2}{\partial\phi^2}\right]+V(r)\right)\Psi_{n\ell m}(r)=E\Psi_{n\ell m}(r),
\label{E17}
\end{equation}
and setting the wave functions $\Psi_{n\ell m}(r)=R_{n\ell}(r)Y_{\ell m}(\theta,\phi)r^{-1}$, we obtain the radial part of the equation as
\begin{equation}
\left[\frac{d^2}{dr^2}+\frac{2\mu}{\hbar^2}\left(E_{n\ell}+\frac{V_o}{1+q\exp{(\frac{r-R}{a})}}\right)-\frac{\ell(\ell+1)}{r^2}\right]R_{n\ell}(r)=0.
\label{E18}
\end{equation}
Because of the total angular momentum centrifugal term, equation (\ref{E18}) cannot be solved analytically for $\ell\neq0$. Therefore, we shall use the Pekeris approximation in order to deal with this centrifugal term and we may express it as follows$\cite{BJ18, BJ19}$
\begin{equation}
U_{cent.}(r)=\frac{1}{r^2}=\frac{1}{R^2\left(1+\frac{x}{R}\right)^2}\cong\frac{1}{R^2}\left(1-2\left(\frac{x}{R}\right)+3\left(\frac{x}{R}\right)^2+\cdots\right),
\label{E19}
\end{equation}
with $x=r-R$. In addition, we may also approximately express it in the following way
\begin{equation}
\tilde{U}=\frac{1}{r^2}\cong\frac{1}{R^2}\left[D_o+\frac{D_1}{1+q\exp{(\nu x)}}+\frac{D_2}{\left(1+q\exp{(\nu x)}\right)^2}\right],
\label{E20}
\end{equation}
where $\nu=1/a$. After expanding (\ref{E20}) in terms of $x$, $x^2$, $x^3$,$\cdots$ and next, comparing with equation (\ref{E19}), we obtain expansion coefficients $D_o$, $D_1$ and $D_2$ as follows:
\begin{eqnarray}
D_o&=&1-\frac{(1+q)^2}{\nu Rq^2}\left(1-\frac{3}{\nu R}\right),\nonumber\\ 
D_1&=&\frac{(1+q)^2}{\nu Rq^2}\left(-1+3q-\frac{6(1+q)}{\nu R}\right),\\
D_2&=&\frac{(1+q)^3}{\nu Rq^2}\left(\frac{1-q}{2}+\frac{3(1+q)}{\nu R}\right).\nonumber
\label{E21}
\end{eqnarray}
Now, inserting the approximation expression (\ref{E20}) into equation (\ref{E18}) and changing the variables $r\rightarrow z$ through the mapping function $z=(1+q\exp{(\nu x)})^{-1}$, equation (\ref{E18}) turns to
\begin{eqnarray}
&&\frac{d^2R_{n\ell}(z)}{dz^2}+\frac{1-2z}{z(1-z)}\frac{dR_{n\ell}(z)}{dz}+\frac{1}{\left[\nu z(1-z)\right]^2}\nonumber\\
&&\times\left[\left(\frac{2\mu E_{n\ell}}{\hbar^2}-\frac{D_o\ell(\ell+1)}{R^2}\right)-z\left(-\frac{2\mu V_o}{\hbar^2}+\frac{D_1\ell(\ell+1)}{R^2}\right)-\frac{D_2z^2\ell(\ell+1)}{R^2}\right]R_{n\ell}(z)=0.
\label{E22}
\end{eqnarray}
Before applying the AIM to this problem, we have to obtain the asymptotic wave functions and then transform equation (\ref{E22}) into a suitable form of the AIM. This can be achieved by the analysis of the asymptotic behaviours at the origin and at infinity. As a result the boundary conditions of the wave functions $R_{n\ell}(z)$ are taken as follows:
\begin{eqnarray}
R_{n\ell}(z)&\rightarrow& 0 \ \ \ \ when\ \ \ \ z\rightarrow 1,\nonumber\\
R_{n\ell}(z)&\rightarrow& 0 \ \ \ \ when\ \ \ \ z\rightarrow 0,
\label{E23}
\end{eqnarray}
thus, one can write the wave functions for this problem as
\begin{equation}
R_{n\ell}(z)=z^\alpha(1-z)^\gamma F_{n\ell}(z),
\label{E24}
\end{equation}
where we have introduced parameters $\alpha$ and $\gamma$ defined by 
\begin{eqnarray}
\alpha&=&\frac{1}{\nu}\left[\frac{D_o\ell(\ell+1)}{R^2}-\frac{2\mu E_{n\ell}}{\hbar^2}\right]^{\frac{1}{2}},\nonumber\\
\gamma&=&\frac{1}{\nu}\left[-\frac{2\mu}{\hbar^2}(V_o+E_{n\ell})+\frac{\ell(\ell+1)}{R^2}(D_o+D_1+D_2)\right]^{\frac{1}{2}},
\label{E25}
\end{eqnarray}
for simplicity. By substituting equation (\ref{E24}) into equation (\ref{E22}), we have the second-order homogeneous differential equation of the form:
\begin{equation}
F_{n\ell}''(z)+\left[\frac{(2\alpha+1)-2z(\alpha+\gamma+1)}{z(1-z)}\right]F_{n\ell}'(z)-\left[\frac{\alpha^2+\gamma^2+\alpha+\gamma+2\alpha\gamma-\frac{D_2\ell(\ell+1)}{\nu^2R^2}}{z(1-z)}\right]F_{n\ell}(z)=0,
\label{E26}
\end{equation}
which is now suitable to an AIM solutions. By comparing this equation with equation (\ref{E1}), we can write the $\lambda_o(z)$ and $s_o(z)$ values and consequently; by means of equation (\ref{E7}), we may derive the $\lambda_k(z)$ and $s_k(z)$ as follows:
\begin{eqnarray}
\lambda_o(z)&=&\frac{2z(\alpha+\gamma+1)-(2\alpha+1)}{z(1-z)},\nonumber\\
s_o(z)&=&\frac{\alpha^2+\gamma^2+\alpha+\gamma+2\alpha\gamma-\frac{D_2\ell(\ell+1)}{\nu^2R^2}}{z(1-z)},\nonumber\\
\lambda_1(z)&=&\frac{2+z\left(2-3\gamma+\gamma^2-\frac{D_2\ell(\ell+1)}{\nu^2R^2}\right)-z^2\left(7\gamma-3\gamma^2-2-\frac{D_2\ell(\ell+1)}{\nu^2R^2}\right)}{z^2(1-z)^2}\\
            &&\frac{[a(z-1)(z(6\gamma-7)-6)]+a^2(4-7z+3z^2)}{z^2(1-z)^2}\nonumber\\
s_1(z)&=&\frac{2\left[(\alpha+\gamma)^2+\alpha+\gamma-\frac{D_2\ell(\ell+1)}{\nu^2R^2}\right]\left[\gamma z-(1+\alpha(1-z))\right]}{z^2(1-z)^2}\nonumber\\
\ldots etc.\nonumber
\label{E27}
\end{eqnarray}
The substitution of the above equations into equation (\ref{E10}), we obtain the first $\delta$ values as
\begin{equation}
\delta_o(z)=\frac{\left[(\alpha+\gamma)^2+(\alpha+\gamma)-\frac{D_2\ell(\ell+1)}{\nu^2R^2}\right]\left[\alpha(3+2\gamma)+\gamma(3+\gamma)-2+\alpha^2-\frac{D_2\ell(\ell+1)}{\nu^2R^2}\right]}{z^2(1-z)^2}.
\label{E28}
\end{equation}
From the root of equation (\ref{E28}), we obtain the first relation between $\alpha$ and $\gamma$ as $\gamma_o+\alpha_o=-\frac{1}{2}-\frac{1}{2}\sqrt{1+4\frac{D_2\ell(\ell+1)}{\nu^2R^2}}$. In a similar fashion, we can obtain other $\delta$ values and consequently establish a relationship between $\alpha_n$ and $\gamma_n$, $n=1, 2, 3,$ $\cdots$ as
\begin{eqnarray}
\delta_1(z)=
\left|
\begin{array}{lr}     
\lambda_2(z)&s_2(z) \\      
  \lambda_{1}(z)&s_{1}(z)
  \end{array}
  \right|=0\ \ \ \ \Rightarrow\ \ \ \ \gamma_1+\alpha_1=-\frac{3}{2}-\frac{1}{2}\sqrt{1+4\frac{D_2\ell(\ell+1)}{\nu^2R^2}}\nonumber\\
  \delta_2(z)=
\left|
\begin{array}{lr}     
\lambda_3(z)&s_3(z) \\      
\lambda_{2}(z)&s_{2}(z)
\end{array}
\right|=0\ \ \ \ \Rightarrow\ \ \ \ \gamma_2+\alpha_2=-\frac{5}{2}-\frac{1}{2}\sqrt{1+4\frac{D_2\ell(\ell+1)}{\nu^2R^2}}\nonumber\\
\ldots etc.
\label{E29}
\end{eqnarray}
The nth term of the above arithmetic progression is found to be
\begin{equation}
\alpha_n+\gamma_n=-\frac{2n+1}{2}-\frac{1}{2}\sqrt{1+4\frac{D_2\ell(\ell+1)}{\nu^2R^2}}.
\label{E30}
\end{equation}
By substituting for $\alpha$ and $\gamma$, we obtain a more explicit expression for the eigenvalues energy as
\begin{equation}
E_{n\ell}=\frac{\hbar^2D_o\ell(\ell+1)}{2\mu R^2}-\frac{\hbar^2}{8\mu a^2}\left[\frac{\left(\frac{1}{2}\sqrt{1+4\frac{D_2a^2\ell(\ell+1)}{R^2}}+\frac{2n+1}{2}\right)^2+a^2\left(\frac{2\mu V_o}{\hbar^2}-\frac{\ell(\ell+1)}{R^2}(D_1+D_2)\right)}{\frac{1}{2}\sqrt{1+4\frac{D_2a^2\ell(\ell+1)}{R^2}}+\frac{2n+1}{2}}\right]^2.
\label{E31}
\end{equation}
Let us now turn to the calculation of the normalized wave functions. By comparing equation (\ref{E26}) with equation (\ref{E13}) we have the following:
\begin{equation}
t=\frac{2\gamma+1}{2},\ \ \ \ b=1,\ \ \ \ N=-1,\ \ \ \ m=\frac{2\alpha-1}{2},\ \ \ \ \sigma=2\alpha+1,\ \ \ \ \rho=2(\alpha-\gamma)+1. 
\label{E32}
\end{equation}
Having determined these parameters, we can easily find the wave functions as
\begin{equation}
F_{n\ell}(z)=(-1)^nC_2\frac{\Gamma(2\alpha+n+1)}{\Gamma{(2\alpha+1)}}\ _2F_1\left(-n, 2(\alpha-\gamma)+1+n; 2\alpha+1; z\right),
\label{E33}
\end{equation}
where $\Gamma$ and $_2F_1$ are the Gamma function and hypergeometric function respectively. By using equations (\ref{E24}) and (\ref{E33}), the total radial wave function can be written as follows:
\begin{equation}
R_{n\ell}(r)=(-1)^nN_{n\ell}\frac{\left[1+q^{-1}\exp{\left(\frac{R-r}{a}\right)}\right]^\gamma}{\left[1+q\exp{\left(\frac{r-R}{a}\right)}\right]^\alpha}\ {_2F_1\left(-n, 2(\alpha-\gamma)+1+n; 2\alpha+1; \left(1+q\exp{\left(\frac{r-R}{a}\right)}\right)^{-1}\right)},
\label{E34}
\end{equation}
where $N_{n\ell}$ is the normalization constant
\newpage
\begin{figure}[h!]
\centering
\includegraphics[height=80mm,width=130mm]{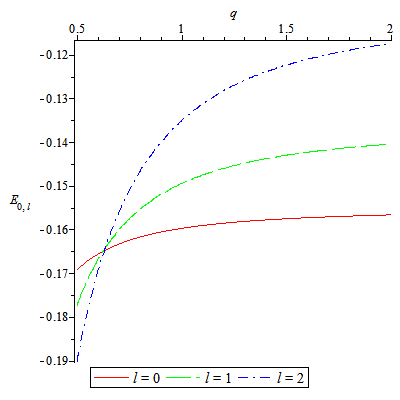}
\caption{\footnotesize The variation of energy spectrum (\ref{E31}) as a function of the deformation parameter. For example we select $\mu=1fm^{-1}$, $V_o=0.3fm^{-1}$, $R=7fm$ and $a=0.65fm$. The radial quantum number is fixed to $n=0$} 
\label{fig1}
\end{figure}
\begin{figure}[h!]
\centering
\includegraphics[height=80mm,width=130mm]{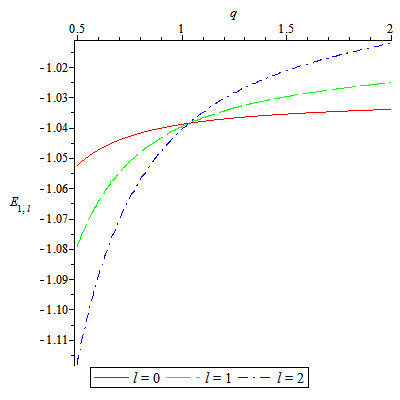}
\caption{\footnotesize The variation of energy spectrum (\ref{E31}) as a function of the deformation parameter. For example we select $\mu=1fm^{-1}$, $V_o=0.3fm^{-1}$, $R=7fm$ and $a=0.65fm$. The radial quantum number is fixed to $n=1$.} 
\label{fig2}
\end{figure}
\section{Results and Conclusion}
In Figures (\ref{fig1}) and (\ref{fig2}) we plot the energy levels $E_{n\ell}$ versus deformation constant $q$ for ground $n=0$ and first exited $n=1$ states, respectively. In Figure (\ref{fig1}), it is seen that when $q\approx0.6$, the orbital quantum numbers $\ell=0, 1, 2,$ will have the same energy eigenvalues, i.e., $E\approx-0.165fm^{-1}$. For the case when $q<0.6$, we noticed that $E_{00}>E_{01}>E_{02}$, whereas when $q>0.6$, then $E_{02}>E_{01}>E_{00}$ (less negative or attractive). In Figure (\ref{fig2}), taking same parameter set, we noticed that when $q\approx1.0$, the three states have energy; $E\approx-1.04fm^{-1}$. In the case when $q<1.0$, $E_{10}>E_{11}>E_{12}$ but when $q>1.0$, $E_{12}>E_{11}>E_{10}$. The system becomes weakly attractive. We have seen that the energy states are sensitive to the deformation constant $q$. There are no physical energy states for $\ell=0,1,2$ when $q\approx0.6$ for $n=0$ and when $q<1.0$ for $n=1$

\begin{figure}[h!]
\centering
\includegraphics[height=75mm,width=130mm]{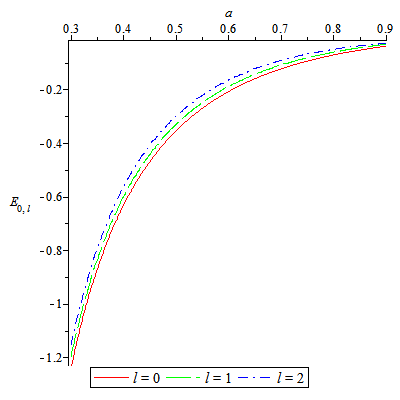}
\caption{\footnotesize The variation of energy spectrum (\ref{E31}) as a function of the diffuseness of the nuclear surface. For example we select $\mu=1fm^{-1}$, $V_o=0.3fm^{-1}$, $R=7fm$ and $q=1.5fm$. The radial quantum number is fixed to $n=0$.} 
\label{fig3}
\end{figure}
\begin{figure}[h!]
\centering
\includegraphics[height=75mm,width=130mm]{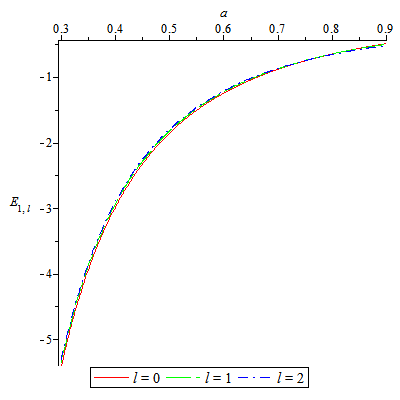}
\caption{\footnotesize The variation of energy spectrum (\ref{E31}) as a function of the diffuseness of the nuclear surface. For example we select $\mu=1fm^{-1}$, $V_o=0.3fm^{-1}$, $R=7fm$ and $q=1.5fm$. The radial quantum number is fixed to $n=1$.} 
\label{fig4}
\end{figure}
In Figures (\ref{fig3}) and (\ref{fig4}), we plot the energy levels $E_{n\ell}$ versus the diffuseness of nuclear surface $a$ for $n=0$ and $n=1$ states, respectively. For $n=0$ case, the energy eigenvalues increasing as $a$ increases, i.e., becoming weakly attractive ($E_{02}>E_{01}>E_{00}$). However, for $n=1$, the case is same as former (i.e. $n=1$) but the three curves overlap each other.

\begin{figure}[h!]
\centering
\includegraphics[height=80mm,width=130mm]{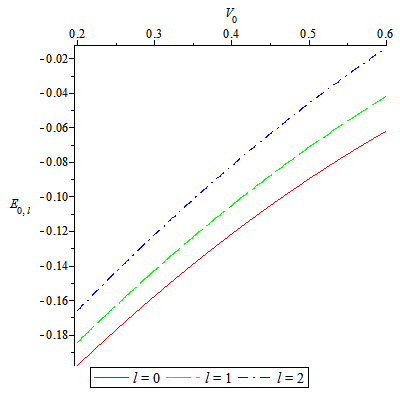}
\caption{\footnotesize The variation of energy spectrum (\ref{E31}) as a function of the potential depth. For example we select $\mu=1fm^{-1}$, $a=0.65fm$, $R=7fm$ and $q=1.5fm$. The radial quantum number is fixed to $n=0$.} 
\label{fig5}
\end{figure}
\begin{figure}[h!]
\centering
\includegraphics[height=80mm,width=130mm]{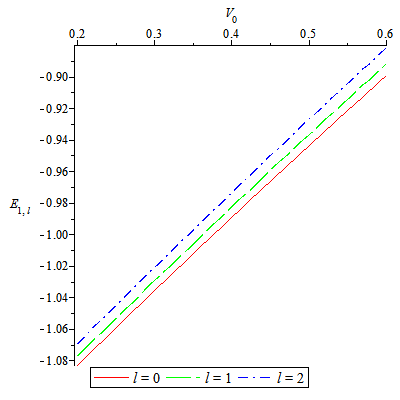}
\caption{\footnotesize The variation of energy spectrum (\ref{E31}) as a function of the potential depth. For example we select $\mu=1fm^{-1}$, $a=0.65fm$ ,$R=0.7fm$ and $q=1.5fm$. The radial quantum number is fixed to $n=1$.} 
\label{fig6}
\end{figure}
In Figures (\ref{fig5}) and (\ref{fig6}), we plot the energy levels $E_{n\ell}$ versus the potential depth potential depth $V_o$ for $n=0$ and $n=1$ states, respectively. For $n=0$ case, we found that there is a substantial change in the energy eigenvalues $E_{02}>E_{01}>E_{00}$,
they increase with increasing $V_o$ and becoming less attractive. However, the same behaviour is seen with slow change in three curves when $n=1$.

\begin{figure}[h!]
\centering
\includegraphics[height=80mm,width=130mm]{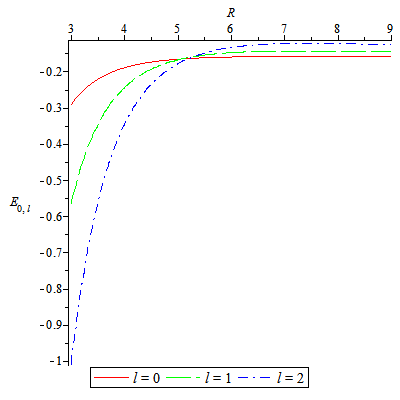}
\caption{\footnotesize The variation of energy spectrum (\ref{E31}) as a function of the potential width. For example we select $\mu=1fm^{-1}$, $V_o=0.3fm^{-1}$, $a=0.65fm$ and $q=1.5$. The radial quantum number is fixed to $n=0$.} 
\label{fig7}
\end{figure}
\begin{figure}[h!]
\centering
\includegraphics[height=80mm,width=130mm]{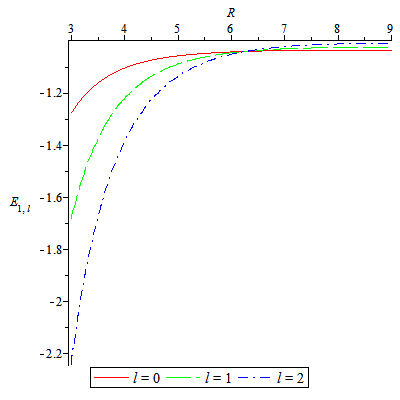}
\caption{\footnotesize The variation of energy spectrum (\ref{E31}) as a function of the potential width. For example we select $\mu=1fm^{-1}$, $V_o=0.3fm^{-1}$, $a=0.65fm$ and $q=1.5$. The radial quantum number is fixed to $n=1$.} 
\label{fig8}
\end{figure}
In Figures (\ref{fig7}) and (\ref{fig8}) we plot the energy levels $E_{n\ell}$   versus the potential width $R$ for the two radial quantum numbers. At $R\approx5.1fm$, the energy is same (coincides). For $R<5.1fm$, $E_{00}>E_{01}>E_{02}$ (less attractive) but when $R>5.1fm$, $E_{02}>E_{01}>E_{00}$. It is obvious that the energy levels are sensitive to potential width $R$.  There are no physical energies when $R<5.1fm$ for $n=0$, and $R<6.2fm$ for $n=1$ . For $E\approx0$, the particles are no longer attracted. However, for $E_{01}$ and $E_{00}\approx-0.2$. When $n=1$, $R\approx6.2fm$, $E_{02}=E_{01}=E_{00}$. For different values of $R$, the behavior is seen as former for $n=0$.

\begin{figure}[h!]
\centering
\includegraphics[height=80mm,width=130mm]{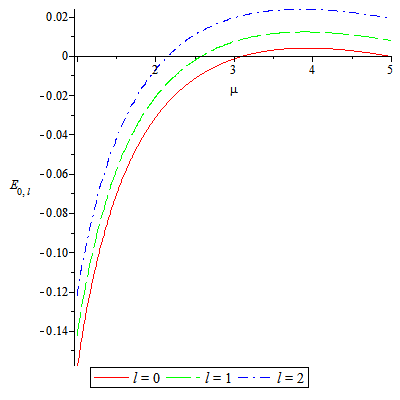}
\caption{\footnotesize The variation of energy spectrum (\ref{E31}) as a function of the particle mass. For example we select $V_o=0.3fm^{-1}$, $R=7fm$, $a=0.65fm$ and $q=1.5$. The radial quantum number is fixed to $n=0$.} 
\label{fig9}
\end{figure}
\begin{figure}[h!]
\centering
\includegraphics[height=80mm,width=130mm]{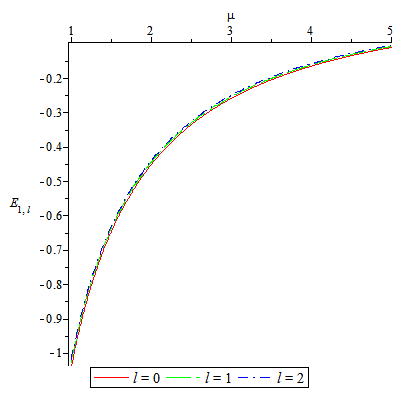}
\caption{\footnotesize The variation of energy spectrum (\ref{E31}) as a function of the particle mass. For example we select $V_o=0.3fm^{-1}$, $R=7fm$, $a=0.65fm$ and $q=1.5$. The radial quantum number is fixed to $n=1$.} 
\label{fig10}
\end{figure}
In Figures (\ref{fig9}) and (\ref{fig10}) we plot the energy levels $E_{n\ell}$ versus the reduced mass $\mu$ for the two radial quantum numbers. In the ground level the energy becomes repulsive (positive) as $\mu$ increases. For example, $E_{02}$ becomes positive for $\mu>2.2fm^{-1}$, $E_{01}>0$ for $\mu>2.5fm^{-1}$ and $E_{00}>0$ for $\mu>3.0fm^{-1}$. This means that the mass has limit. When mass increases, then we have no attractive particles and hence no energy spectrum. However, in the excited level, the energies of the three states are close to each other and the system remains attractive when the reduced mass increases. 

We have seen that the approximately analytical bound states solutions of the $\ell-$wave Schr$\ddot{o}$dinger equation for the nuclear deformed Woods-Saxon potential can be solved by proper approximation to the centrifugal term within the framework of the AIM. By using the AIM, closed analytical forms for the energy eigenvalues are obtained and the corresponding wave functions have been presented in terms of hypergeometric functions.

The method presented in this paper is an elegant and powerful technique. If there are analytically solvable potentials, it provides the closed forms for the eigenvalues and the corresponding eigenfunctions. However, the case if the solution is not available, the eigenvalues are obtained by using an iterative approach $\cite{BJ21, BJ22, BJ23}$.


\begin{thebibliography}{99}
\bibitem{BJ1} R. D. Woods and D. S. Saxon, Phys. Rev. {\bf95} (1954) 577.
\bibitem{BJ2} W. S. C. Williams, Nuclear and Particle Physics, Clarendon, Oxford, 1996.
\bibitem{BJ3} F. Garcia et al., Eur. Phys. J. A {\bf6} (1999) 49.
\bibitem{BJ4} V. Goldberg et al., Phys. Rev. C {\bf69} (2004) 031302.
\bibitem{BJ5} A. Syntfelt et al., Eur. Phys. J. A {\bf20} (2004) 359.
\bibitem{BJ6} A. Diaz-Torres and W. Scheid, Nucl. Phys. A {\bf757} (2005) 373.
\bibitem{BJ7} J. Y. Guo and Q. Sheng, Phys. Lett. A {\bf338} (2005) 90.
\bibitem{BJ8} V. A.Chepurnov and P. E. Nemirovsky, Nucl. Phys. {\bf49} (1963) 90.
\bibitem{BJ9} R. R. Chasman and B. D. Wilkins, Phys. Lett. B {\bf149} (1984) 433. 
\bibitem{BJ10} J. Dudek and T. Wemer, J. Phys. G: nuclear Phys. {\bf4} (1978) 1543.
\bibitem{BJ11} S. Fl$\ddot{u}$gge, Practical Quantum Mechanics, Springer-Verlag, Berlin, 1974.
\bibitem{BJ12} V. H. Badalov, H. I. Ahmadov, S. V. Badalov, Int. J. Mod. Phys. E {\bf18} (2010) 1463.
\bibitem{BJ13} V. H. Badalov, H. I. Ahmadov, S. V. Badalov, Int. J. Mod. Phys. E {\bf18} (2009) 631.
\bibitem{BJ14} S. M. Ikhdair and R. Sever, Cent. Eur. J. Phys. {\bf8} (2010) 652; Int. J. Mod. Phys. A {\bf25} (2010) 3941.
\bibitem{BJ15} M. Hamzavi and S. M. Ikhdair, Few-Body Syst, DOI 10.1007/s00601-012-0452-9 (2012).
\bibitem{BJS1} S.M. Ikhdair and R. Sever, Int. J. Theor. Phys. {\bf46} (2007) 1643.
\bibitem{BJ16} H. Ciftci, R. L. Hall and N. Saad, J. Phys. A: Math Gen. {\bf36} (2003) 11807. 
\bibitem{BJ17} H. Ciftci, R. L. Hall and N. Saad, Phys. Lett. A: {\bf340} (2005) 388.
\bibitem{BJS2} S.M. Ikhdair and R. Sever, Ann. Phys. (Leipzig) {\bf16} (2007) 218.
\bibitem{BJ18} L. I. Schiff, \textit{Quantum Mechanics 3rd edn.} (McGraw-Hill Book Co., New York, 1968).
\bibitem{BJ19} L. D. Landau and E.M. Lifshitz, \textit{Quantum Mechanics, Non-relativistic Theory, 3rd edn.} (Pergamon, New York, 1977).
\bibitem{BJ20} V. H. Badalov, H. I. Ahmadov, and S. V. Badalov, arXiv:0912.3890v2 math-ph.
\bibitem{BJ21} T. Barakat, J. Phys. A: Math. Gen. {\bf36} (2006) 823.
\bibitem{BJ22} F. M. Fernandez, J. Phys. A: Math. Gen. {\bf37} (2004) 6173.
\bibitem{BJ23} T. Barakat, Phys. Lett. A {\bf344} (2005) 411.
\end{thebibliography}
\end{document}